# HARMONIC CENTRALIZATION OF SOME GRAPH FAMILIES

**Jose Mari E. Ortega\* and Rolito G. Eballe**

Mathematics Unit
Philippine Science High School Southern Mindanao Campus
Davao City 8000, Philippines
e-mail: josemari.ortega@smc.pshs.edu.ph

Department of Mathematics
Central Mindanao University
Musuan, Bukidnon, Philippines
e-mail: rgeballe@cmu.edu.ph

## Abstract

Centrality describes the importance of nodes in a graph and is modeled by various measures. Its global analogue, called centralization, is a general formula for calculating a graph-level centrality score based on the node-level centrality measure. The latter enables us to compare graphs based on the extent to which the connections of a given network are concentrated on a single vertex or group of vertices. One of the measures of centrality in social network analysis is harmonic centrality. It sums the inverse of the geodesic distances of each node to other nodes where it is 0 if there is no path from one node to another, with the sum normalized by dividing it by $m - 1$, where $m$ is the number of nodes of the graph. In this paper, we present some results regarding the harmonic centralization of some important families of graphs with the hope that formulas generated herein will be of use when one determines the harmonic centralization of more complex graphs.

## 1. Introduction

In graph theory and social network analysis, the notion of centrality is based on the importance of the nodes in a graph. In 1979, Freeman [5] expounded on the concept of centrality being an important attribute of social networks where its value may relate to other important properties of said networks. Gómez [6] presented a number of centrality measures detailing their structural characteristics and specific set-ups, each with its strengths and qualities depending on the major concerns for which those were originally conceptualized.

A more recent centrality measure is the harmonic centrality proposed by Marchiori and Latora [11] in 2000 and then later by Dekker [2] in 2005 and by Rochat [14] in 2009. Designed to take the sum of the reciprocals of the distances of vertex $u$ from each of the other vertices in a graph, it was invented to solve the problem of dealing with disconnected graphs by equating the required reciprocal to 0 if there is no path from the concerned node to another. The harmonic centrality then normalizes the sum by dividing it by $m - 1$, where $m$ is the number of nodes in the graph. Ortega and Eballe [13] determined the harmonic centrality of the vertices of some graph families. For related works with closeness and betweenness centrality of some graph families, see [3] and [15].

While centrality quantifies the importance of a node in a graph, centralization (also known as group centralization or Freeman centralization) is a general formula for calculating a graph-level centrality score based on the node-level centrality measure. It sums up the differences in the centrality of each vertex from the vertex with the highest centrality, where this sum is then normalized by the most centralized graph, a star graph with the same number of the vertices. An immediate advantage of this graph-level centralization is that it allows us to compare how centralized graphs are based on the extent to which the connections of a given graph are concentrated on a single vertex or group of vertices. Butts [1] investigated on the degree centralization on some graphs while Unnithan et al. [15] studied betweenness centralization of some classes of graphs. Krnc et al. also published a number of papers on centralization, see [7-10].

In this paper, we aim to determine the harmonic centralization of some important classes of graphs. In particular, we will consider the path graph $P_m$, cycle graph $C_m$, fan graph $F_m$, wheel graph $W_m$, complete bipartite graph $K_{m,n}$, ladder graph $L_m$, crown graph $Cr_m$, prism graph $Y_m$, book graph $B_m$, helm graph $H_m$, and complete split graph $CS(n, k)$. Please be it noted that all graphs considered here are in the context of being finite, simple, and undirected.

## 2. Preliminaries and Some General Properties

For formality, we provide the definitions of the main parameters discussed in this paper and then generate some of basic properties. General descriptions of the special families of graphs are provided herein as well.

**Definition 2.1** (Harmonic centrality of a vertex in a graph)**.** Let $G = (V(G), E(G))$ be a nontrivial graph of order $m$. If $u \in V(G)$, then the *harmonic centrality* of $u$ is given by the expression

$$\mathcal{H}_G(u) = \frac{\mathcal{R}_G(u)}{m-1},$$

where $\mathcal{R}_G(u) = \sum_{x \in V(G)\setminus\{u\}} \frac{1}{d(x, u)}$ with $d(x, u)$ being the geodesic distance between vertices $x$ and $u$ such that $\frac{1}{d(x, u)} = 0$ in case there is no path from $x$ to $u$ in $G$.

**Example 2.2.** Figure 1 shows a caterpillar graph $G$ with $u \in V(G)$, where we have $d(x_1, u) = 2$, $d(x_2, u) = 1$, $d(x_3, u) = 2$, $d(x_4, u) = 2$, $d(x_5, u) = 1$, $d(x_6, u) = 2$, $d(x_7, u) = 1$, $d(x_8, u) = 2$, and $d(x_9, u) = 2$.

Thus, $\mathcal{R}_G(u) = \sum_{x=1}^{9} \frac{1}{d(x, u)} = \frac{1}{2} + 1 + \frac{1}{2} + \frac{1}{2} + 1 + \frac{1}{2} + 1 + \frac{1}{2} + \frac{1}{2} + 6$. Since $m = 10$, $\mathcal{H}_G(u) = \frac{\mathcal{R}_G(u)}{m-1} = \frac{6}{9} = \frac{2}{3}$.

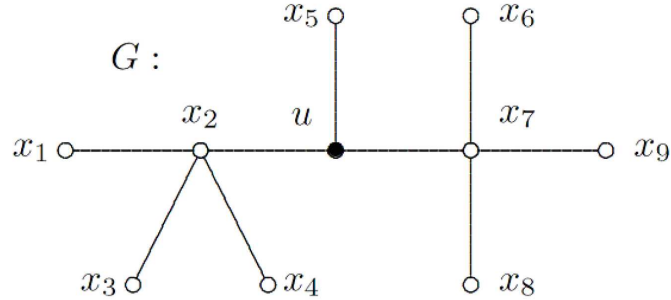

**Figure 1.** A caterpillar graph $G$ with $u \in V(G)$, where $\mathcal{H}_G(u) = \frac{2}{3}$.

In a similar fashion, we can compute $\mathcal{H}_G(x_1) = \frac{47}{108}$, $\mathcal{H}_G(x_2) = \frac{2}{3}$, $\mathcal{H}_G(x_3) = \frac{47}{108}$, $\mathcal{H}_G(x_4) = \frac{47}{108}$, $\mathcal{H}_G(x_5) = \frac{4}{9}$, $\mathcal{H}_G(x_6) = \frac{47}{108}$, $\mathcal{H}_G(x_7) = \frac{2}{3}$, $\mathcal{H}_G(x_8) = \frac{47}{108}$, and $\mathcal{H}_G(x_9) = \frac{47}{108}$.

The *harmonic centralization* of a graph $G$ of order $m \geq 2$ is given by the Freeman centralization formula

$$C_{\mathcal{H}}(G) = \frac{\sum_{i=1}^{m}(\mathcal{H}_{G\,\max}(u) - \mathcal{H}_G(u_i))}{\max \sum_{i=1}^{m}(\mathcal{H}_{G^*\,\max}(u^*) - \mathcal{H}_{G^*}(u_i^*))}, \quad (1)$$

where $\mathcal{H}_{G\,\max}(u)$ is the largest-valued harmonic centrality of vertex $u$ in graph $G$, while the denominator is evaluated over all graphs of order $m$ [5]. But according to Rochat [14], the denominator of the formula in (1) is attained by the star graph $K_{1,\,m-1}$ of order $m$. In this case, if $u_0$ is the vertex of the star graph $K_{1,\,m-1}$ with the highest degree, then the denominator of the Freeman centralization formula (1) reduces to

$$\sum_{u_i \in V(K_{1,\,m-1})} (\mathcal{H}_{K_{1,\,m-1}}(u_0) - \mathcal{H}_{K_{1,\,m-1}}(u_i)) = (m-1)\left(1 - \frac{m}{2m-2}\right) = \frac{m-2}{2}.$$

**Definition 2.3.** For any graph $G$ of order $m > 2$, the *harmonic centralization* of $G$ is given by the formula

$$C_{\mathcal{H}}(G) = \frac{\sum_{i=1}^{m}(\mathcal{H}_{G\,\max}(u) - \mathcal{H}_G(u_i))}{\frac{m-2}{2}},$$

where $\mathcal{H}_{G\,\max}(u)$ is the largest-valued harmonic centrality of a vertex $u$ in $G$.

**Example 2.4.** Consider the graph $G$ given in Figure 1. The harmonic centralization of this graph is computed by finding the sum of the differences of the vertex with the maximum harmonic centrality with all other vertices and then normalizing it by the most centralized graph, the star graph of the same order. So by Definition 2.3,

$$C_{\mathcal{H}}(G) = \frac{\sum_{i=1}^{m}(\mathcal{H}_{G\,\max}(u) - \mathcal{H}_G(u_i))}{\frac{m-2}{2}} = \frac{\sum_{i=1}^{m}(\mathcal{H}_G(u) - \mathcal{H}_G(u_i))}{\frac{10-2}{2}}$$

$$= \frac{\sum_{i=1}^{10}\frac{2}{3} - \sum_{i=1}^{10}\mathcal{H}_G(u_i)}{4}$$

$$= \frac{\frac{20}{3} - \left[\frac{2}{3} + \frac{47}{108} + \frac{2}{3} + \frac{47}{108} + \frac{47}{108} + \frac{4}{9} + \frac{47}{108} + \frac{2}{3} + \frac{47}{108} + \frac{47}{108}\right]}{4}$$

$$= \frac{29}{72}.$$

**Proposition 2.5.** *Let G be a graph of order* $m > 2$. *Then* $0 \leq C_{\mathcal{H}}(G) \leq 1$.

From the Freeman harmonic centralization formula given in (1), we immediately have

$$\sum_{i=1}^{m}(\mathcal{H}_{G\,\max}(u) - \mathcal{H}_G(u_i)) \leq \max \sum_{i=1}^{m}(\mathcal{H}_{G^*\,\max}(u^*) - \mathcal{H}_{G^*}(u_i^*)),$$

where $\max \sum_{i=1}^{m}(\mathcal{H}_{G^*\,\max}(u^*) - \mathcal{H}_{G^*}(u_i^*)) = \frac{m-2}{2} > 0,$ since $m > 2$ implying that $C_{\mathcal{H}}(G) \leq 1$. In addition, since $\sum_{i=1}^{m}(\mathcal{H}_{G\,\max}(u) - \mathcal{H}_G(u_i)) \geq 0$, we also have $C_{\mathcal{H}}(G) \geq 0$. □

**Corollary 2.6.** *The harmonic centralization of the complete graph* $K_m$ *of order* $m > 2$ *is zero.*

**Proof.** This follows from the fact that all the vertices of the complete graph $K_m$, $m > 2$ have a uniform harmonic centrality of 1. □

**Definition 2.7** (Harmonic number $H_n$). The *nth harmonic number* $H_n$ is the sum of the reciprocals of the first $n$ natural numbers; that is,

$$H_n = 1 + \frac{1}{2} + \frac{1}{3} + \cdots + \frac{1}{n} = \sum_{k=1}^{n} \frac{1}{k}.$$

Again, in this paper, the harmonic centralization of some special graphs such as the path graph $P_m$, cycle graph $C_m$, fan graph $F_m$, wheel graph $W_m$, complete bipartite graph $K_{m,n}$, ladder graph $L_m$, crown graph $Cr_m$, star graph $S_m$, prism graph $Y_m$, book graph $B_m$, helm graph $H_m$, and complete split graph $CS(n, k)$ shall be derived. Most of these families of graphs were considered in [12].

Recall that the path graph $P_m$ of order $m$ is a graph with distinct vertices $a_1, a_2, \ldots, a_m$ and edges $a_1 a_2, a_2 a_3, \ldots, a_{m-1} a_m$, while the cycle graph $C_m$ of order $m \geq 3$ is a graph with distinct vertices $a_1, a_2, \ldots, a_m$ and edges $a_1 a_2, a_2 a_3, \ldots, a_{m-1} a_m, a_m a_1$. On the other hand, the fan graph $F_m$ of order $m + 1$, where $m \geq 3$, is a graph formed by adjoining one vertex $u_0$ to each vertex of path $P_m = [u_1, u_2, \ldots, u_m]$. Figure 2 shows the skeletal diagrams of the path, cycle and fan graphs.

The wheel graph $W_m$ of order $m + 1$, $m > 3$, is a graph formed by adjoining one vertex $u_0$ to each vertex of cycle $C_m = [u_1, u_2, \ldots, u_m, u_1]$. The complete bipartite graph $K_{m,n}$, where both $m, n \geq 2$, has

$$V(K_{m,n}) = \{u_1, u_2, \ldots, u_m\} \cup \{v_1, v_2, \ldots, v_n\},$$

$$E(K_{m,n}) = \{u_i v_j \mid 1 \leq i \leq m, 1 \leq j \leq n\}.$$

On the other hand, the ladder graph $L_m$ of order $2m$ is a graph formed as the Cartesian product of a path graph $P_m = [u_1, u_2, \ldots, u_m]$ with the path graph $P_2 = [v_1, v_2]$. Figure 3 shows the skeletal diagrams for a wheel graph, a complete bipartite graph, and a ladder graph.

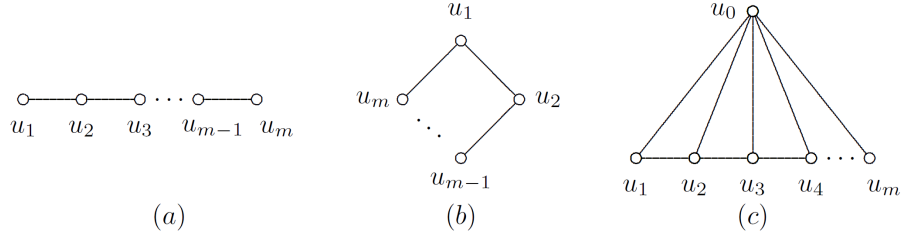

**Figure 2.** (a) Path graph $P_m$; (b) cycle graph $C_m$; and (c) fan graph $F_m$.

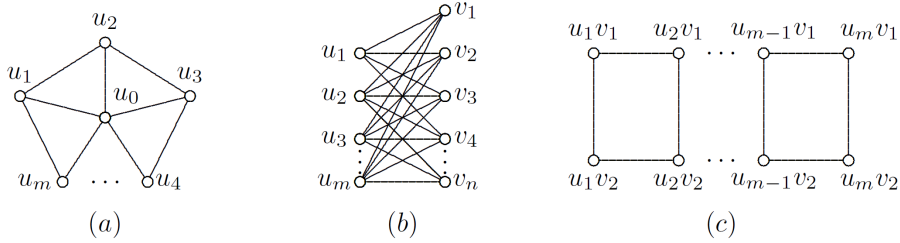

**Figure 3.** (a) Wheel graph $W_m$; (b) complete bipartite graph $K_{m,n}$; and (c) ladder graph $L_m$.

The crown graph $Cr_m$ of order $2m$ is the graph with $V(Cr_m) = \{u_1, u_2, ..., u_m\} \cup \{v_1, v_2, ..., v_m\}$ and whose edges are formed by adjoining $u_i$ to $v_j$ whenever $i \neq j$. The prism graph $Y_m$ of order $2m$, with $m \geq 3$, is a graph formed as the Cartesian product of a cycle graph $C_m = [u_1, u_2, ..., u_m, u_1]$ with the path graph $P_2 = [v_1, v_2]$. The star graph $S_m$ of order $m+1$, $m > 1$, is a graph formed by adjoining $m$ isolated vertices $u_i$, $1 \leq i \leq m$, to a single vertex $u_0$. Figure 4 shows the skeletal diagrams for the crown, prism and star graphs.

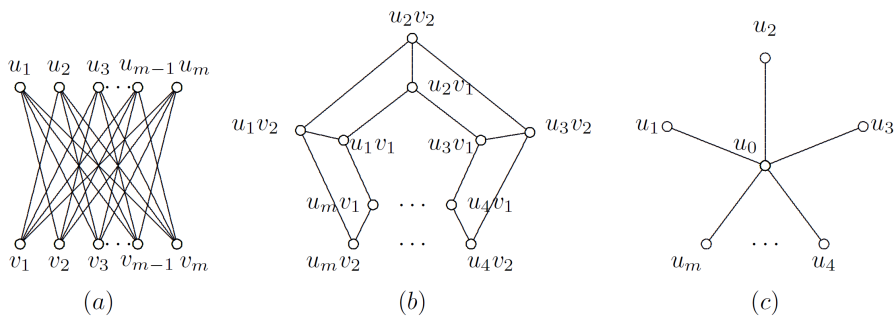

**Figure 4.** (a) Crown graph $Cr_m$; (b) prism graph $Y_m$; and (c) star graph $S_m$.

The book graph $B_m$ of order $2(m+1)$ is formed as the Cartesian product of a star graph $S_m$ (with center vertex $u_0$) with the path graph $P_2 = [v_1, v_2]$. The helm graph $H_m$ of order $2m+1$, $m \geq 3$, is obtained by adjoining a pendant vertex at each node of the $m$-ordered cycle of the wheel graph $W_m$, with the vertices $V(H_m) = [u_0, u_1, ..., u_m] \cup [v_1, v_2, ..., v_m]$. The complete split graph $CS(n, k)$ of order $n + k$ is formed by adjoining two subsets $A = \{a_1, a_2, ..., a_n\}$ and $B = \{b_1, b_2, ..., b_n\}$ such that $\langle A \rangle$ is a clique and $B$ is an independent set, where every vertex in $B$ is adjacent to every vertex in the clique $\langle A \rangle$. Figure 5 shows the skeletal diagrams of a book graph, a helm graph, and a complete split graph.

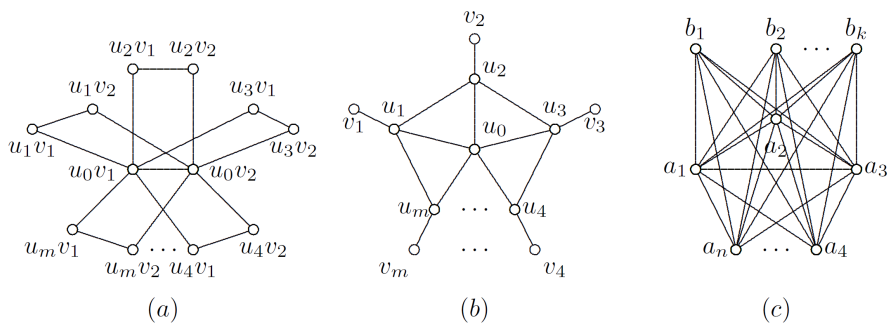

**Figure 5.** (a) Book graph $B_m$; (b) helm graph $H_m$; and (c) complete split graph $CS(n, k)$.

### 3. Harmonic Centralization of Some Special Families of Graphs

In this section, we determine the harmonic centralization of some special graph families, namely, the path graph $P_m$, cycle graph $C_m$, fan graph $F_m$, wheel graph $W_m$, complete bipartite graph $K_{m,n}$, ladder graph $L_m$, crown graph $Cr_m$, prism graph $Y_m$, star graph $S_m$, book graph $B_m$, helm graph $H_m$, and complete split graph $CS(n, k)$.

**Theorem 3.1.** *For the path graph $P_m$ of order m, where $m > 2$, its harmonic centralization is given by*

$$C_{\mathcal{H}}(P_m) = \begin{cases} \dfrac{4}{(m-1)(m-2)}\left[(m-1)H_{\frac{m-1}{2}} - H_{m-1} - \sum_{i=2}^{\frac{m-1}{2}}(H_{i-1} + H_{m-i})\right], & \text{if } m \text{ is odd}, \\ \dfrac{4}{(m-1)(m-2)}\left[\dfrac{m-2}{m} + (m-2)H_{\frac{m-2}{2}} - H_{m-1} - \sum_{i=2}^{\frac{m-2}{2}}(H_{i-1} + H_{m-i})\right], & \text{if } m \text{ is even}. \end{cases}$$

**Proof.** If the order $m$ of a path graph $P_m = [u_1, u_2, ..., u_m]$ is odd, then the vertex $u_{\frac{m+1}{2}}$ will have the maximum harmonic centrality of $\dfrac{2}{m-1}H_{\frac{m-1}{2}}$. To find the graph centralization, we add the differences of the maximum harmonic centrality and the harmonic centrality of the nodes for $i = 1$, $i = m$, and for $1 < i < m$. Thus,

$C_{\mathcal{H}}(P_m)$

$$= \frac{1}{\frac{m-2}{2}} \left[ 2\left( \frac{2}{m-1} H_{\frac{m-1}{2}} - \frac{2}{m-1} H_{m-1} \right) \right.$$

$$\left. + \frac{2}{m-1}\left( 2\left( \frac{m-3}{2} \right) \right) H_{\frac{m-1}{2}} - \sum_{i=2}^{\frac{m-1}{2}} (H_{i-1} + H_{m-i}) \right]$$

$$= \frac{4}{(m-1)(m-2)} \left[ 2H_{\frac{m-1}{2}} + (m-3)H_{\frac{m-1}{2}} - H_{m-1} - \sum_{i=2}^{\frac{m-1}{2}} (H_{i-1} + H_{m-i}) \right]$$

$$= \frac{4}{(m-1)(m-2)} \left[ (m-1)H_{\frac{m-1}{2}} - H_{m-1} - \sum_{i=2}^{\frac{m-1}{2}} (H_{i-1} + H_{m-i}) \right].$$

On the other hand, if $m$ is even, then the vertices $u_{\frac{m}{2}}$ and $u_{\frac{m+2}{2}}$ will have the maximum harmonic centrality of $\frac{1}{m-1}\left( \frac{2}{m} + 2H_{\frac{m-2}{2}} \right)$. With the same procedure as above, we add up the differences of the maximum harmonic centrality and the harmonic centrality of the nodes for $i = 1$, $i = m$, and for $1 < i < m$. So,

$C_{\mathcal{H}}(P_m)$

$$= \frac{1}{\frac{m-2}{2}}\left[\frac{2}{m-1}\left(\frac{2}{m} + 2H_{\frac{m-2}{2}} - H_{m-1}\right)\right.$$

$$\left. + \frac{2}{m-1}\left(\frac{m-4}{2}\left(\frac{2}{m} + 2H_{\frac{m-2}{2}} - \sum_{i=2}^{\frac{m-2}{2}}(H_{i-1} + H_{m-i})\right)\right)\right]$$

$$= \frac{4}{(m-1)(m-2)}\left[\frac{2}{m} + 2H_{\frac{m-2}{2}} - H_{m-1} + \frac{m-4}{m}\right.$$

$$\left. + (m-4)H_{\frac{m-2}{2}} - \sum_{i=2}^{\frac{m-2}{2}}(H_{i-1} + H_{m-i})\right]$$

$$= \frac{4}{(m-1)(m-2)}\left[\frac{m-2}{m} + (m-2)H_{\frac{m-2}{2}} - H_{m-1} - \sum_{i=2}^{\frac{m-2}{2}}(H_{i-1} + H_{m-i})\right].$$

□

**Theorem 3.2.** *For the cycle graph $C_m$ of order m, with $m \geq 3$, the harmonic centralization is zero.*

**Proof.** From [13], $\mathcal{H}_{C_m}(u_i) = \frac{2}{m-1}(H_{\frac{m-1}{2}})$ if $m$ is odd and $\mathcal{H}_{C_m}(u_i)$

$= \frac{2}{m-1}\left(H_{\frac{m-1}{2}} + \frac{1}{m}\right)$ if $m$ is even. In either case, each node in a cycle graph will have the same harmonic centrality value. Therefore, the resulting harmonic centralization is zero. □

**Theorem 3.3.** *For the fan graph $F_m$ of order $m+1$, where $m \geq 3$, the harmonic centralization is given by*

$$C_{\mathcal{H}}(F_m) = \frac{m-2}{m}.$$

**Proof.** The maximum harmonic centrality of a fan graph is 1 at the vertex $u_0$, with $\mathcal{H}_{F_m}(u_i) = \frac{m+2}{2m}$ for $i = 1, m$ and $\mathcal{H}_{F_m}(u_i) = \frac{m+3}{2m}$ for $1 < i < m$, so the harmonic centralization of a fan graph is given by

$$C_{\mathcal{H}}(F_m) = \left(\frac{1}{\frac{m-1}{2}}\right)\left[2\left(1 - \frac{m+2}{2m}\right) + (m-2)\left(1 - \frac{m+3}{2m}\right)\right]$$

$$= \left(\frac{2}{m-1}\right)\left(\frac{m-2}{m} + \frac{(m-2)(m-3)}{2m}\right)$$

$$= \left(\frac{2}{m-1}\right)\left(\frac{2m - 4 + m^2 - 5m + 6}{2m}\right) = \frac{m^2 - 3m + 2}{m(m-1)}$$

$$= \frac{m-2}{m}. \qquad \square$$

**Theorem 3.4.** *For the wheel graph $W_m$ of order $m+1$, the harmonic centralization is given by*

$$C_{\mathcal{H}}(W_m) = \frac{m-3}{m-1}.$$

**Proof.** The harmonic centrality is 1 for the center vertex $u_0$ and $\frac{m+3}{2m}$ otherwise, so the harmonic centralization of a wheel graph is given by

$$C_{\mathcal{H}}(W_m) = \left(\frac{1}{\frac{m-1}{2}}\right)\left(m\left(1 - \frac{m+3}{2m}\right)\right) = \left(\frac{2}{m-1}\right)\left(\frac{m(m-3)}{2m}\right) = \frac{m-3}{m-1}. \qquad \square$$

**Theorem 3.5.** *For the complete bipartite graph $K_{m,n}$ of order $m + n$, where both $m, n > 1$, the harmonic centralization is given by*

$$C_{\mathcal{H}}(K_{m,n}) = \begin{cases} 0, & \text{if } m = n, \\ \dfrac{m(m-n)}{(m+n-2)(m+n-1)}, & \text{if } m > n, \\ \dfrac{n(n-m)}{(m+n-2)(m+n-1)}, & \text{if } m < n. \end{cases}$$

**Proof.** The harmonic centralization of a complete bipartite graph $K_{m,n}$ is determined by the values of $m$ and $n$, with the following cases:

**Case 1.** If $m = n$, then $\mathcal{H}_{K_{m,n}}(u_i) = \mathcal{H}_{K_{m,n}}(v_j)$. Thus, $C_{\mathcal{H}}(K_{m,n}) = 0$.

**Case 2.** If $m > n$, then the vertices with the maximum harmonic centrality will be the $v_j$'s. Thus,

$$C_{\mathcal{H}}(K_{m,n}) = \left(\dfrac{1}{\dfrac{m+n-2}{2}}\right)\left[m\left(\dfrac{2m+n-1}{2(m+n-1)} - \dfrac{m+2n-1}{2(m+n-1)}\right)\right]$$

$$= \left(\dfrac{2}{m+n-2}\right)\left(\dfrac{m(m-n)}{2(m+n-1)}\right)$$

$$= \dfrac{m(m-n)}{(m+n-2)(m+n-1)}.$$

**Case 3.** If $m < n$, then the vertices with the maximum harmonic centrality will be the $u_i$'s. Thus,

$$C_{\mathcal{H}}(K_{m,n}) = \left(\frac{1}{\frac{m+n-2}{2}}\right)\left[n\left(\frac{m+2n-1}{2(m+n-1)} - \frac{2m+n-1}{2(m+n-1)}\right)\right]$$

$$= \left(\frac{2}{m+n-2}\right)\left(\frac{n(n-m)}{2(m+n-1)}\right)$$

$$= \frac{n(n-m)}{(m+n-2)(m+n-1)}. \qquad \square$$

**Theorem 3.6.** *For the ladder graph $L_m$ of order $2m$, where $m \geq 3$, the harmonic centralization is given by*

$$C_{\mathcal{H}}(L_m) = \begin{cases} \frac{4}{(m-1)(2m-1)}\left[2(m-1)H_{\frac{m-1}{2}} - 2H_{m-1} + \frac{2(m-1)}{m+1} - \frac{m-1}{2}\right. \\ \qquad \left. - \frac{1}{m} - \sum_{i=2}^{\frac{m-1}{2}}\left(2H_{i-1} + 2H_{m-i} + \frac{1-i}{i} + \frac{1}{m-i+1}\right)\right], & \text{if } m \text{ is odd,} \\[2ex] \frac{2}{(2m-1)(m-1)}\left[4(m-2)H_{\frac{m}{2}} - 4H_{m-1} - \frac{m^2-2}{m} + \frac{2m-4}{m+2}\right. \\ \qquad \left. - 2\sum_{i=2}^{\frac{m-2}{2}}\left(2H_{i-1} + 2H_{m-i} + \frac{1-i}{i} + \frac{1}{m-i+1}\right)\right]. & \text{if } m \text{ is even.} \end{cases}$$

**Proof.** In a ladder graph of order $2m$, if $m$ is odd, then the vertices $(u_{\frac{m+1}{2}}, v_1)$ and $(u_{\frac{m+1}{2}}, v_2)$ will have the maximum harmonic centrality of $\frac{4}{2m-1}\left(H_{\frac{m-1}{2}} + \frac{1}{m+1} - \frac{1}{4}\right)$. To find the ladder graph centralization, we add up the differences of the maximum harmonic centrality and the harmonic centrality of the nodes for $i = 1$, $i = m$, and $1 < i < m$ for all $1 \leq j \leq 2$ from [13]. So,

$$C_{\mathcal{H}}(L_m) = \frac{1}{\frac{2m-2}{2}} \left[ \left( \frac{4(2(m-1))}{2m-1} \left( H_{\frac{m-1}{2}} + \frac{1}{m+1} - \frac{1}{4} \right) \right) \right.$$

$$- \frac{4}{2m-1} \left( 2H_{m-1} + \frac{1}{m} \right)$$

$$\left. - \frac{4}{2m-1} \sum_{i=2}^{\frac{m-1}{2}} \left( 2H_{i-1} + 2H_{m-i} + \frac{1-i}{i} + \frac{1}{m-i+1} \right) \right]$$

$$= \frac{4}{(m-1)(2m-1)} \left[ 2(m-1)H_{\frac{m-1}{2}} - 2H_{m-1} + \frac{2(m-1)}{m+1} \right.$$

$$\left. - \frac{m-1}{2} - \frac{1}{m} - \sum_{i=2}^{\frac{m-1}{2}} \left( 2H_{i-1} + 2H_{m-i} + \frac{1-i}{i} + \frac{1}{m-i+1} \right) \right].$$

On the other hand, if $m$ is even in a ladder graph of order $2m$, then the vertices $(u_{\frac{m}{2}}, v_1)$, $(u_{\frac{m}{2}}, v_2)$, $(u_{\frac{m+2}{2}}, v_1)$ and $(u_{\frac{m+2}{2}}, v_2)$ will have the maximum harmonic centrality of $\frac{1}{2m-1} \left( 4H_{\frac{m}{2}} - \frac{m+2}{m} + \frac{2}{m+2} \right)$. Then we find the graph centralization by adding the differences of the maximum harmonic centrality and the harmonic centrality of the nodes for $i = 1$, $i = m$, and for $1 < i < m$. So,

$$C_{\mathcal{H}}(L_m) = \frac{1}{\frac{2m-2}{2}} \left[ \frac{4}{2m-1}\left(4H_{\frac{m-1}{2}} - 2H_{m-1} - \frac{m+3}{m} + \frac{2}{m+2}\right) \right.$$

$$+ \frac{2(m-4)}{2m-1}\left(4H_{\frac{m}{2}} - \frac{m+2}{m} + \frac{2}{m+2}\right)$$

$$\left. - \frac{4}{2m-1} \sum_{i=2}^{\frac{m-2}{2}} \left(2H_{i-1} + 2H_{m-i} + \frac{1}{i} + \frac{1}{m-i+1} - 1\right) \right]$$

$$= \frac{2}{(m-1)(2m-1)}$$

$$\cdot \left[ 8H_{\frac{m}{2}} + 4(m-4)H_{\frac{m}{2}} - 4H_{m-1} - \frac{2(m+3) - (m+2)(m-4)}{m} \right.$$

$$\left. + \frac{4 + 2(m-4)}{m+2} - 2\sum_{i=2}^{\frac{m-2}{2}}\left(2H_{i-1} + 2H_{m-i} + \frac{1-i}{i} + \frac{1}{m-i+1}\right) \right]$$

$$= \frac{1}{(2m-1)(m-1)}$$

$$\cdot \left[ 4(m-2)H_{\frac{m}{2}} - 4H_{m-1} - \frac{(m^2-2)}{m} + \frac{2m-4}{m+2} \right.$$

$$\left. - 2\sum_{i=2}^{\frac{m-2}{2}}\left(2H_{i-1} + 2H_{m-i} + \frac{1-i}{i} + \frac{1}{m-i+1}\right) \right]. \qquad \square$$

**Theorem 3.7.** *For the crown graph $Cr_m$ of order $2m$, where $m \geq 3$, the harmonic centralization is zero.*

**Proof.** This follows from the fact that all the vertices of a crown graph $Cr_m$ will have the same harmonic centralities of $\frac{9m-7}{12m-6}$. Therefore, the corresponding harmonic centralization of a crown graph is zero. □

**Theorem 3.8.** *For the prism graph $Y_m$ of order $2m$, where $m \geq 3$, the harmonic centralization is zero.*

**Proof.** Each vertex of the prism graph $Y_m$ will have the same harmonic centralities of $\mathcal{H}_{Y_m}(u_i) = \frac{2}{m-1}\left(4H_{\frac{m-1}{2}} + \frac{3-m}{m+1}\right)$ if $m$ is odd and $\mathcal{H}_{Y_m}(u_i) = \frac{2}{m-1}\left(4H_{\frac{m-1}{2}} + \frac{2}{m+2} - \frac{m+2}{m}\right)$ if $m$ is even. Therefore, this results to the harmonic centralization of zero. □

**Theorem 3.9.** *For the book graph $B_m$ of order $2(m+1)$, where $m \geq 3$, the harmonic centralization is given by*

$$C_{\mathcal{H}}(B_m) = \frac{4(m-1)}{3(2m+1)}.$$

**Proof.** From [13], the vertices with the maximum harmonic centrality will be $(u_0, v_1)$ and $(u_0, v_2)$, so

$$C_{\mathcal{H}}(B_m) = \left(\frac{1}{\frac{2(m+1)-2}{2}}\right)\left[2m\left(\frac{3m+2}{4m+2} - \frac{5(m+2)}{6(2m+1)}\right)\right]$$

$$= \left(\frac{1}{m}\right)\left[2m\left(\frac{4(m-1)(2m+1)}{6(2m+1)(2m+1)}\right)\right]$$

$$= \frac{4(m-1)}{3(2m+1)}. \qquad \square$$

**Theorem 3.10.** *For the helm graph $H_m$ of order $2m + 1$, where $m \geq 3$, the harmonic centralization is given by*

$$C_{\mathcal{H}}(H_m) = \begin{cases} \dfrac{2}{5}, & \text{if } m = 3, \\ \dfrac{19m - 47}{12(2m - 1)}, & \text{if } m > 3. \end{cases}$$

**Proof.** In finding the harmonic centralization of a helm graph, there will be two cases. If $m = 3$, then the vertices with the maximum harmonic centrality will be the $u_i$'s. Thus,

$$C_{\mathcal{H}}(H_m) = \left(\dfrac{1}{\dfrac{2m+1-2}{2}}\right)\left[\left(\dfrac{5m+15}{12m} - \dfrac{3}{4}\right) + 3\left(\dfrac{5m+15}{12m} - \dfrac{7m+17}{24m}\right)\right]$$

$$= \left(\dfrac{2}{5}\right)\left[\left(\dfrac{30}{36} - \dfrac{3}{4}\right) + 3\left(\dfrac{30}{36} - \dfrac{38}{72}\right)\right] = \left(\dfrac{2}{5}\right)\left(\dfrac{1}{12} + \dfrac{33}{36}\right) = \dfrac{2}{5}.$$

On the other hand, if $m > 3$, then the vertex $u_0$ will have the maximum harmonic centrality. Thus,

$$C_{\mathcal{H}}(H_m) = \left(\dfrac{1}{\dfrac{2m+1-2}{2}}\right)\left[m\left(\dfrac{3}{4} - \dfrac{5m+15}{12m}\right) + m\left(\dfrac{3}{4} - \dfrac{7m+17}{24m}\right)\right]$$

$$= \left(\dfrac{2m}{2m-1}\right)\left(\dfrac{3}{2} - \dfrac{10m+30}{24m} - \dfrac{7m+17}{24m}\right)$$

$$= \left(\dfrac{2m}{2m-1}\right)\left(\dfrac{36m - 10m - 30 - 7m - 17}{24m}\right)$$

$$= \dfrac{19m - 47}{12(2m-1)}. \qquad \square$$

**Theorem 3.11.** *For the complete split graph $CS(n, k)$ of order $n + k$, where $n \geq 2$ and $k \geq 1$, the harmonic centralization is given by*

$$C_{\mathcal{H}}(CS(n, k)) = \frac{k(k-1)}{(n+k-1)(n+k-2)}.$$

**Proof.** The harmonic centrality will be 1 for $u_i \in A$ and $\dfrac{n + \frac{1}{2}(k-1)}{n+k-1}$ for $u_i \in B$. Thus,

$$C_{\mathcal{H}}(CS(n, k)) = \left(\frac{1}{\frac{n+k-2}{2}}\right)\left[n(1-1) + k\left(1 - \frac{n + \frac{1}{2}(k-1)}{n+k-1}\right)\right]$$

$$= \left(\frac{2}{n+k-2}\right)\left[k\left(\frac{(n+k-1) - \left(n + \frac{1}{2}(k-1)\right)}{n+k-1}\right)\right]$$

$$= \frac{k(k-1)}{(n+k-2)(n+k-1)}. \qquad \square$$